# Gravity as the Second-Order Relativistic-Manifestation of Electrostatic-Force

**R.C.Gupta**
Professor
Institute of Engineering & Technology (I.E.T.)
Lucknow, India
rcg_iet@hotmail.com

**ABSTRACT**

It is well known that magnetic force between two current carrying conductors is a relativistic manifestation of net electrostatic force between relatively moving electrons & protons of the two wires. On similar grounds but with more accurate considerations i.e., of second order relativistic considerations, it is shown here that gravitational-force between two bodies is in-fact due to relativistic manifestation of net electrostatic force between protons & revolving electrons in atoms of the two bodies. As special-relativity provides a bridge between electricity and magnetism, in somewhat similar way (as described in the paper) it is shown that once again it is the special-relativity which provides the much sought-after link between electrostatic force and gravitational force. If magnetism is the first-order (velocity-dependant) effect of moving electrons in conductor, gravitation is the second-order (acceleration-dependant) effect of revolving electrons in atoms. The new proposed gravitation-theory seems to be well *in accordance* with gravity behavior i.e., it correctly predicts that gravity is a long-range attractive-force proportional to masses & follows inverse-square-law and is transmitted with the velocity of light.

## 1. Introduction

Gravity is the most common force around us. But it is surprising that, although there was Newton's formula for gravitation, there was no explanation for 'why gravity is there' till Einstein's [1-4] explanation (1915). Alternative theories of gravity have been reported occasionally, specially on net, such as Kaluza-Klein theory [4-6] and others [7-11], though appealing but not with much acceptance/success. Einstein, through general-relativity, explains that gravity is not a force but appears to be so due to curvature of 4-dimensional space-time. Einstein's general-relativity [1-4] is built on Riemannian curved geometry. It appears as if the gravity is disguised as space-time curvature in tensor notations. Einstein's general-relativity, however, has attained grand success.

Gravitational force (Newton's formula) is similar to Electrostatic force (Coulomb's formula) in many ways. Both follow inverse square law and are long range force. Thus it is tempting to find a clue of gravitational force into electrostatic force. The author proposes such a theory in this paper and claims that 'gravitation *is* due to second-order relativistic-manifestation of electrostatic-force'.

Out of the 5 forces namely: electrostatic, magnetic, gravitational, weak-nuclear and strong-nuclear forces; the electrostatic & magnetic forces are already united (by Faraday/Maxwell) as electro-magnetic force [12]. The electromagnetic force and weak nuclear force were further unified through QED, Feynman rules, symmetry, group theory, gauge theory & renormalization etc.[13] ( by Glashow, Weinberg, Salam) as electro-weak force [13]. On similar lines with quantum-chromodynamics (QCD) the strong force is further considered unified leading to Standard-Model / GUT [13-18]. The standard model has been extremely successful; but problem is with the unification of gravitation as gravity seems to be quite 'refractive' (difficult) toward unification [15-18]. The superstring/M(brane)- theory [15-20] and its variant supergravity/quantum-gravity, currently, are the leading candidates for final unification; it however talks more of the mathematics of much higher dimensions. The present paper, could interestingly provide a possible simple link of gravitation to electromagnetic force; it is shown that as if gravitation is a 'second-order effect' of electrostatic force as magnetism is the 'first-order effect' of electrostatic force, *or* in other words as if gravity is a second-order effect of magnetism between atoms.

It is well known that special-relativity is the 'bridge' between electricity and magnetism [12,21]. In the present paper it is shown that 'how special-relativity is able to provide further link between electrostatic force and gravitational force'. A brief review is given (in the following section - 2) however, to show how the relativity provides a bridge between electricity & magnetism, to illuminate the concept for further utilizing this approach (with little refinement/accuracy taking the relativistic velocity addition into account) subsequently (in sections - 3 & 4) for linking of electrostatic force to gravitational force.

## 2. Magnetism as First-order Relativistic-Manifestation of Electrostatic-Force

Very few people appreciate that magnetism is an effect of electricity via relativity. In literature & books it is explained in several ways. Faraday & Maxwell Equations relate electricity & magnetism mathematically. Simple good explanations are found in some books [12,21], the 'simplified' approach [21] is briefly described in the next paragraph to illuminate the concept for extending it further in section-3 for explaining gravitation.

Consider two parallel conductors I and II carrying current in same direction. Let the electrons and protons in conductor-I be $A_1$ & $P_1$, and that in conductor-II be $A_2$ & $P_2$. The electrons $A_1$ see $A_2$ at rest and $P_2$ moving with velocity v in opposite direction and thus there is a length contraction by an amount ½ $v^2/c^2 = \beta^2/2$ for spacing between protons, or in other words more number of protons appear in the conductor-II. Similarly, to proton $P_1$ more electrons appear in conductor-II. This leads to a net attraction between the two conductors proportional to $v^2/c^2$ which otherwise normally is considered to be due to magnetic effect (which in fact is first-order relativistic-manifestation of electrostatic-force).

## 3. The New Theory of Gravitation :
### Gravitation as Second-Order Relativistic-Manifestation of Electrostatic-Force

It is well known, as explained above, that magnetism is a relativistic-effect of electrostatic force. On similar lines the concept is further extended to show that even gravity too is a relativistic-effect of electrostatic force. This new theory of gravity is proposed & explained in this section. The emphasis, however, is on the concept & qualitative aspects rather than on an accurate quantitative analysis. For simplicity and to make it more illustrative - a simple model is described as following.

Consider two atoms one each in two bodies I & II. For clarity of explanation consider 'Helium'-like atoms. The two electrons are $A_1$ & $B_1$ and the two protons are grouped as $P_1$ in body-I, and that in body-II are $A_2$ & $B_2$ and $P_2$ (Fig.1, neutrons omitted are not shown). For simplicity it is assumed that electrons revolve in circular-type orbits and let their orbits, though in two bodies, be in same Y-Z plane. Consider both electrons in the two orbits revolving clockwise. Furthermore, the position of electrons are considered to be such that (Fig.1) $A_1$ & $A_2$ are moving in parallel(same) direction and that $A_1$ & $B_2$ are moving in antiparallel(opposite) direction each with velocity v. With this arrangement, the net attraction or repulsion factor between atoms in I and II can be found considering length contraction and are tabulated in the important Table 1, *without* relativistic velocity addition in the middle column and *with* relativistic velocity addition in the last column.

The author calls 'length-contraction *without* relativistic velocity addition' as 'first-order relativistic effect' and 'length-contraction *with* relativistic velocity addition', as 'second-order relativistic effect'. It is to be noted (see Total in the table) that only the simple length contraction which is able to link electricity and magnetism, here fails to provide any resultant attraction between two atoms of the two bodies *unless* relativistic velocity addition is taken into account before the length contraction. Thus (in view of Fig.1 and Table 1), it can be said that 'gravitation *is* due to second order relativistic effect of electrostatic force between atoms of the two bodies'. Even if it is considered that the electrons in body-I revolving clockwise and that in body-II revolving anti-clockwise, the net result (last row of Total in Table 1) would be same. So, the second-order effect always causes net *attraction* between two matter-bodies irrespective of direction of electron rotation.

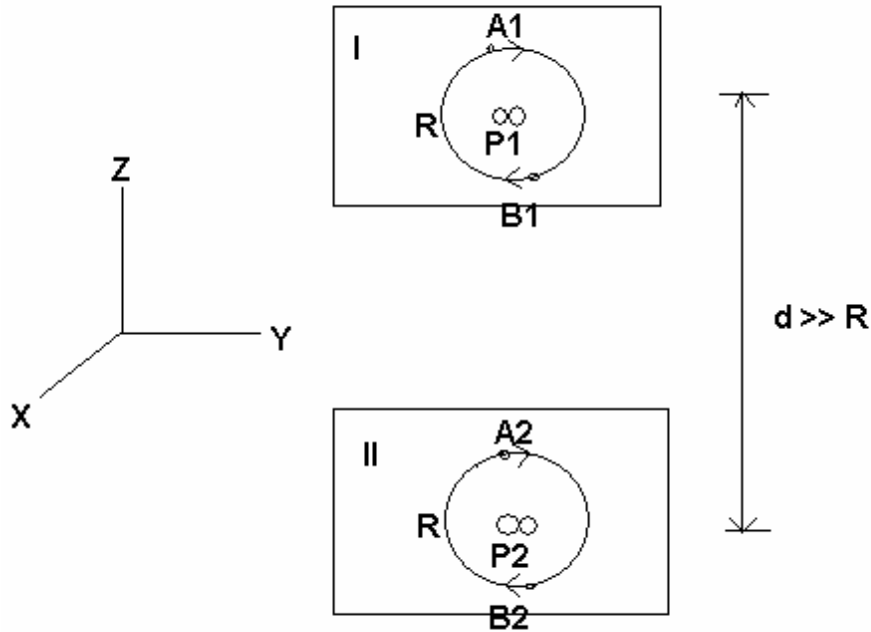

FIG 1   ATOMS IN TWO BODIES

**Table.1**  Net Attraction (+) or Repulsion (-) Factor between Electrons & Protons of the Two Atoms
in the Two Matter Bodies I & II  due to Length Contraction of Relativity

| Observations | Net attraction(+) or repulsion(-) Factor due to length contraction | | | |
|---|---|---|---|---|
| | *without* relativistic velocity addition (first-order effect) | | *with* relativistic velocity addition (second-order effect) | |
| $A_1$ as observer sees $A_2$ | 0 | = 0 | 0 | = 0 |
| $A_1$ as observer sees $P_2$ | $+2 \times \tfrac{1}{2} v^2/c^2$ | $= +\beta^2$ | $+2 \times \tfrac{1}{2} v^2/c^2$ | $= +\beta^2$ |
| $A_1$ as observer sees $B_2$ | $-\tfrac{1}{2}(2v)^2/c^2$ | $= -2\beta^2$ | $-\tfrac{1}{2}(2v)^2/c^2 \cdot \{1/(1+v^2/c^2)^2\}$ | $\cong -2\beta^2(1-2\beta^2)$ |
| $B_1$ as observer sees $A_2$ | $-\tfrac{1}{2}(2v)^2/c^2$ | $= -2\beta^2$ | $-\tfrac{1}{2}(2v)^2/c^2 \cdot \{1/(1+v^2/c^2)^2\}$ | $\cong -2\beta^2(1-2\beta^2)$ |
| $B_1$ as observer sees $P_2$ | $+2 \times \tfrac{1}{2} v^2/c^2$ | $= +\beta^2$ | $+2 \times \tfrac{1}{2} v^2/c^2$ | $= +\beta^2$ |
| $B_1$ as observer sees $B_2$ | 0 | = 0 | 0 | = 0 |
| $P_1$ as observer sees $A_2$ | $+2 \times \tfrac{1}{2} v^2/c^2$ | $= +\beta^2$ | $+2 \times \tfrac{1}{2} v^2/c^2$ | $= +\beta^2$ |
| $P_1$ as observer sees $P_2$ | 0 | = 0 | 0 | = 0 |
| $P_1$ as observer sees $B_2$ | $+2 \times \tfrac{1}{2} v^2/c^2$ | $= +\beta^2$ | $+2 \times \tfrac{1}{2} v^2/c^2$ | $= +\beta^2$ |
| **atom-I as observer sees atom-II** | **Total** | **= 0** | **Total** | $\cong +8\beta^4$ |

The clue to link gravitational and electrostatic forces lies in the *relativistic velocity addition* due to inherent *asymmetry* hidden in it. In fact the velocity addition must be made relativistically before the length contraction (or net attraction or repulsion factor) estimation. For example, $A_1$, as observer sees, $B_2$ moving not with velocity $2v$ but with velocity $2v/(1 + v^2/c^2)$. It is seen that thus the resultant net attraction factor (total of last column in Table1) is *not zero* but equal to $8\beta^4 = 8(v/c)^4$ due to relativistic velocity added length contraction in Y-direction (Fig.1). For 'Hydrogen'-like atom this factor will be less, and can similarly be shown to be only $2\beta^4$.

If both bodies-atoms are considered to lie in X-Z or X-Y plane (Fig.1), similar results could be found. However, if the orbit plane in body-I lies in Y-Z plane but that in body-II lies in X-Y plane, the velocity of electrons in body I and II being perpendicular, the length-contraction(attraction-factor) will be zero. Hence it is concluded that total net attraction factor between two bodies (containing many atoms having different orbit orientations) would be very much less than $\beta^4$. Moreover, the likelihood of electrons positions as shown in Fig.1 is extremely-small estimated of the order of $(r/R)^4$, $r$ & $R$ being the electron-radius & orbit-radius respectively; thus the total net resultant attraction factor is further drastically reduced by this factor (as follows).

## 4. Derivation of Newton's gravitational formula from Coulomb's electrostatic force formula and Estimation of Gravitational Constant G

Probability of the electron(of radius $r$) at point $A_1$ on the electron's orbital cloud shell (of radius $R$) is approx. $(\pi r^2)/(4\pi R^2)$. Similar probability is for the electron $A_2$. Probability for such combination as $A_1$ & $A_2$ as shown in Fig.1 is thus estimated as of the order of $(r/R)^4$. The net attraction factor due to second order relativistic effect for atoms (Fig.1) shown in the Table1 is of the order of $(v/c)^4$. Thus, considering the extremely less likelihood of electrons positions as depicted in Fig 1, the total net resultant attraction factor is proportional to $(v/c)^4 \cdot (r/R)^4$. In absence of a more generalized/accurate analysis, it is better to write the total net resultant attraction factor as $f = K \cdot (v/c)^{n1} \cdot (r/R)^{n2}$. Also, only a small fraction of electrons (say, 2%) in one body may be compatible(position & orbit wise), in view of relativistic length contraction, with similar small fraction of electrons in the other body; thus $K \ll 1$ of the order of $4 \times 10^{-4}$. It may, however, be noted here that in present paper the emphasis is on the qualitative-concepts rather than quantitative-estimation; so even a rough estimate with a simple model will suffice the purpose.

Now consider a body has mass $m_1$ kg $= 10^3 m_1$ gram, and atomic weight A gram & atomic number Z; thus number of moles $= 10^3 m_1/A$, number of atoms $= (10^3 m_1/A) \cdot A_N$ where Avogadro-number $A_N = 6.02 \times 10^{23}$. So, number of protons or electrons $= (m_1 10^3/A) \cdot A_N \cdot Z$. Using mass number $A = N + Z$ and taking number of neutrons $N \cong Z$ approximately; $A \cong 2 \cdot Z$. Thus number of protons or electrons present would approximately be $\frac{1}{2} A_N \cdot 10^3 m_1$ in body-I and $\frac{1}{2} A_N \cdot 10^3 m_2$ in body-II.

So attraction force between the two bodies I & II, considering the number of electrons & protons and the total net resultant attraction factor as mentioned above, is the gravitational attraction between two bodies I & II resulting in due to second-order relativistic-manifestation of electrostatic-force and is given by:

$$F = \{1/(4\pi\varepsilon)\} \cdot q_1 \cdot [q_2 \cdot f]/d^2 \tag{1.a}$$

$$= \{1/(4\pi\varepsilon)\} \cdot (\tfrac{1}{2} A_N \cdot 10^3 m_1 \cdot e) \cdot [(\tfrac{1}{2} A_N \cdot 10^3 m_2 \cdot e) \cdot \{K \cdot (v/c)^{n1} \cdot (r/R)^{n2}\}]/d^2 \tag{1.b}$$

$$= G \cdot m_1 \cdot m_2/d^2 \tag{1.c}$$

Thus, the value of Gravitational-constant G can theoretically be estimated (approximately) from:

$$G = \{(\tfrac{1}{2} A_N 10^3)^2 \cdot e^2/(4\pi\varepsilon)\} \cdot \{K \cdot (v/c)^{n1} \cdot (r/R)^{n2}\} \tag{2}$$

Taking $\{1/(4\pi\varepsilon)\} = 9\times10^9$, $e = 1.6\times10^{-19}$, $A_N = 6.02 \times 10^{23}$, average $v/c \cong 10^{-2}$ & $r/R \cong 10^{-6}$ and selecting/adjusting values of $n_1 = 4$, $n_2 = 4$, $K \cong 4\times10^{-4}$, the value of G is roughly estimated as $8\times10^{-11}$ which is of the same order of magnitude of the known value of $6.7\times10^{-11}$ in SI units.

## 5. Predictions of the New Gravitation Theory

### (A) Predictions which are well in accordance with Gravity

This new theory of gravitation claims that '*gravitation is due to second-order relativistic-manifestation of electrostatic-force*' on similar lines as magnetism is known to be relativistic effect of electricity. Thus gravitational force is an aspect of electromagnetic force. The predictions of this new theory that are well *in accordance* are that : 'Gravitational force between the two matter bodies -

   (i)    is attractive,
   (ii)   is proportional to their masses,
   (iii)  follows inverse square law,
   (iv)   is a long range force as that of the Coulomb's force, but
   (v)    is much weaker in strength, and
   (vi)   the gravitational-signal travels as that for electric-force with speed of light.

### (B) New predictions for Gravity for test / re-interpretation

Following new predictions could be thus made from this new gravitational theory, which can possibly be tested / re-interpreted by scientists -

(i) Gravitational constant G is not a universal constant. It should depend on the *materials* (atoms) of the attracting bodies for which average electron-velocity v and orbit-radius R could be different for different materials. Also, G should depend on permittivity $\varepsilon$ of the *medium*. The dependence could possibly be averaged-out and counter-balanced / masked by other effects, so difficult to verify.

(ii) (a) The force between two matter bodies (I and II) is attractive. But the force between one matter-body (I) and another antimatter-body (II) would be not-attracting but it would be *repulsive* as per this new proposed theory of gravitation (as per Fig.1 in Table1 the Total would be *then* $-8\beta^4$, negative sign would mean *repulsion*). Force between two antimatter bodies will again be attractive. It is interesting to note that although charged particle & antiparticle attract each other, but matter (atoms) & antimatter (anti-atoms) would *repel* each other.

(b) If antimatter (such as anti-hydrogen) is made to collide with high velocity onto the matter, the high velocity can possibly overcome the repulsion between matter & antimatter and ultimately should annihilate each other producing short bursts of energy.

(iii) (a) According to Einstein's general-relativity, the *outer* space-time is 'curved' around material-body as if causing gravity. Although Einstein's theory has passed several tests, but the real test of space-time curvature is on-the-way at Stanford-University [22] under the able guidance of Dr. Francis Everitt with NASA & Lockheed-Martin collaboration. The presently proposed theory of gravitation claims that gravitation *is* second order effect of special-relativity on electrostatic force between atoms and all this comes because electrons revolve in 'closed' thus 'curved' orbital-path *inside* the material-body. According to Einstein - the space-time curvature is *outside* the body whereas as per the new theory - the orbit-curvature is *inside* the body. String-theory [14-18] also talks about *inside curl-up* of extra-dimensions but in a different context & size.

(b) Probably, Einstein's general-relativity-based theory and the author's special-relativity-based (second-order atomic-magnetism) theory of gravitation, both altogether different, may still come to be compatible (*outside* curvature of space-time may possibly be imprint of *inside* curvature of electron's orbit through gravity); possibly some modification and/or re-interpretation of either or both may be needed or some fundamental change in the concepts have to be made. For example one such fundamental change, in the concept of *mass*, has been proposed by the author in another paper [23] whereas it is suggested that in general the mass M is *quaternion* (scalar + vector) quantity: the scalar-part establishes the particle-aspects and the vector-part governs the wave-behavior; which explains the differently-looking wave & particle and its duality, and also in a way makes wide-differing subjects 'Special-relativity' & 'Quantum-mechanics' compatible.

## 5. Discussions

The author means by 'first-order' and 'second-order' relativistic effects as follows:- if velocity addition is done non-relativistically and length contraction of relativity is used then it is considered as first-order relativity effect, whereas if velocity addition is done relativistically and length contraction of relativity is used then it is considered as *second*-order (relativity being used at *two*-stages) relativistic effect (see Table 1). It is concluded that 'gravitation *is* due to second-order relativistic-manifestation of electrostatic-force resulting in due to *revolving* motion of electrons in atoms of the two bodies, in the same way as magnetism is the first order relativistic manifestation of electrostatic force resulting in due to *linear* motion of electrons in two conductors'. In nutshell it can be said that: 'if magnetism is the first-order (velocity-dependant) effect, gravitation is the second-order (as if acceleration-dependant) effect'; *or* in other words, 'gravity is the second-order-relativistic atomic-magnetism due to revolving-electrons'. The acceleration-dependence means being function of centripetal-acceleration $v^2/R$ of the orbiting electrons.

There exists well known [21,24] 'van der Wall' force (attractive) between molecules and also between atoms even without any consideration of length-contraction or relativity, this may be said as if 'zeroth-order' relativistic effect, meaning with 'no' relativistic effect at all. In fact, the van der Wall force is also manifestation of electrostatic force, but due to charge-distribution. This attractive force varies effectively with 'inverse of seventh power' of separation distance d and thus is almost truly zero at large distances; it is a short-range force responsible for several common phenomena such as condensation of gases, freezing of liquid to solid, surface tension, viscosity, friction etc. But it can not lead to gravitational effect as it is short range force and varies sharply as $1/d^7$. Any other variation (other than inverse square), if any, such as $1/d^3$ too can not be responsible for gravitation.

It is true that, most of small differences don't make much of a difference, but if the small difference pops up at crucial point then it can make all the difference [25]. The second-order relativistic manifestation is very small but it adds up to make all the difference by giving birth of gravity from atomic-electricity. Einstein himself said [25] that 'Books on physics are full of complicated formulae. But thoughts and ideas, not formulae, are the beginning of every physical theory'. So, emphasis in the present paper is on the concept; and the simple formulation seems good enough to begin with. The gravitational-theory presented here that '*gravitational force is second order relativistic manifestation of electrostatic force*' seems reasonably correct as it very-rightly predicts that gravitation is long range inverse square law force, having features quite similar to electrostatic force but with much smaller strength.

The theory developed in the paper indicates that *gravitation* is a weak *electro-magnetic* effect of atoms of the mass, so it seems somewhat obvious that light (the *electro-magnetic* wave) should show deflection and red-shift in presence of gravity due to *electro-magnetic* 'interactions'. Moreover, as gravity is caused due to velocity (v) of orbiting-electrons of atoms of the mass; and if the mass (planet) itself is moving, its own-velocity (u) addition would change the gravitational force a bit to cause a possible precession (advance) of perihelion of the planet. Detailed mechanism / understanding, explanation & calculations of these, not that easy, need to be worked out, however. The author's theory suggests that gravitation is electro-magnetic phenomenon and its electromagnetic-interaction with the electro-magnetic-wave can cause light to move in curved path in the medium near the star-mass; which otherwise is understood in general-relativity as if gravity is no-force but the space-time itself is curved around the mass and test-particle moves in such geodesic. Also, since gravity-signal is due to electromagnetic-effect it travels with velocity of light, and the exchange-particle would have properties similar to photon (zero rest mass) but spin may be different than 1 (could be 2, as that of graviton) possibly due to electromagnetic 'interaction' of weak *photon-pair* signals (as if coming from two positions ( as A1 & B1 and A2 & B2 in Fig.1) of electrons in atomic-orbits). Gravitation is in-fact electro-magnetic phenomenon; you may call it - attraction/repulsion, electromagnetic-interaction, space-time curvature / warping or geodesic / light-bending or by whatever-name.

The biggest impact of new understanding of gravitation would be on understanding of cosmos / universe [26,27]. The known things such as planet, star, galaxy & black-hole and mysterious things such as dark-matter & dark-energy may have to be looked-at in new-lights from different-perspective. It may be noted that it seems possible [28] to explain some general-relativity phenomena/tests, including black-hole, *without* general-relativity. The proposed gravitation-theory (as second-order atomic-magnetism) can grow & develop subsequently, and could eventually prove to be a step forward for better understanding of physics, astro-physics & cosmology.

### 6. Conclusions

Gravity is the most common force around us but unfortunately there is no simple & clear explanation for it. A simple, clear and alternative explanation is proposed in the present paper. Contrast to Einstein's general-relativity-based explanation of gravity, the author's explanation is based on special-relativity. The new proposed gravito-electro-magnetic theory propounds that '*gravitational-force is due to second-order relativistic-manifestation of electrostatic-force*' on somewhat similar lines as magnetism is relativistic-effect of electricity. The proposed theory is well *in accordance* with gravity-behavior (Newton's formula) having *features* quite similar to its origin electrostatic-force (Coulomb's formula), and it thus provides a possible link towards unification of the forces. The new gravitation-theory tells that gravitation is in-fact the second-order electro-magnetic phenomenon and the gravitational-*attraction* and / or space-time-*curvature / geodesic* are due to the electromagnetic-*interaction*.

### Acknowledgement

The author dedicates this fundamental research work to his parents Late Mrs. Vidyawati Devi and Late Mr. B.P.Gupta. The author is also thankful to Dr. Abdus Salam of ICTP Trieste, Dr. Vinod Johri, Prof. Emeritus, Dr. B. Das, Asstt.Professor of Lucknow University and Dr. M.S.Kalara, Professor, I.I.T. Kanpur for their encouragement & discussions. Thanks are also due to Veena , Ruchi, Sanjay, Chhavi, Sanjiv and Shefali for their support & assistance. The author thanks Dr.V.P.Gautam, Dr. Sanjay Mishra, Dr. Ravi Sinha, Prof. H.N.Gupta, Prof. K.K.Srivastava and Er. Anil Gupta for their advice & help. AICTE and IET are acknowledged with thanks for the R&D and TEQIP grants & facilities.

### References/Bibliography

---